\documentclass[apj,iop]{emulateapj}
\usepackage{multirow}
\usepackage{graphicx}
\usepackage{amsmath, amsthm, amssymb}
\usepackage{natbib}
\usepackage{hyperref}

\shorttitle{Radial migration and the vertical structure of galaxy disks}
\shortauthors{Vera-Ciro, D'Onghia \& Navarro}
\graphicspath{{figs/}}

\hypersetup{
   bookmarks=true,              
   unicode=false,               
   pdftoolbar=true,             
   pdfmenubar=true,             
   pdffitwindow=false,          
   pdfstartview={FitH},         
   pdftitle={Rad Migration},    
   pdfauthor={Vera-Ciro et al.},
   pdfnewwindow=true,           
   colorlinks=true,             
   linkcolor=blue,              
   citecolor=blue,              
   filecolor=blue,              
   urlcolor=blue                
 }

\newcommand{\diskone}{\texttt{LD}}
\newcommand{\disktwo}{\texttt{HD}}
\newcommand{\diskthree}{\texttt{HD-MW}}

\begin{document}

\title{The imprint of radial migration on the vertical structure of galaxy disks}

\author{Carlos Vera-Ciro\altaffilmark{1,2}, Elena D'Onghia\altaffilmark{1,4}, 
Julio F. Navarro\altaffilmark{3,5}}

\affil{\textsuperscript{1} Department of Astronomy, University of Wisconsin, 475
  N. Charter Street, Madison, WI 53076, USA} 
\email{e-mail:cvera@udem.edu.co}

\affil{\textsuperscript{2} Departamento de Ciencias B\'asicas, Universidad de Medell\'in, Cra 87 N
30-65, Medell\'in, Colombia} 
\affil{\textsuperscript{3} Department of Physics and Astronomy, 
University of Victoria, Victoria, BC, V8P 5C2, Canada}

\begin{abstract}
  We use numerical simulations to examine the effects of radial
  migration on the vertical structure of galaxy disks. The simulations
  follow three exponential disks of different mass but similar
  circular velocity, radial scalelength, and (constant) scale
  height. The disks develop different non-axisymmetric patterns,
  ranging from feeble, long-lived multiple arms to strong,
  rapidly-evolving few-armed spirals. These fluctuations induce radial
  migration through secular changes in the angular momentum of disk
  particles, mixing the disk radially and blurring pre-existing
  gradients. Migration affects primarily stars with small vertical
  excursions, regardless of spiral pattern. This ``provenance bias''
  largely determines the vertical structure of migrating stars: inward
  migrators thin down as they move in, whereas outward migrators do
  not thicken up but rather preserve the disk scale height at
  destination.  Migrators of equal birth radius thus develop a strong
  scale-height gradient, not by flaring out as commonly assumed, but
  by thinning down as they spread inward. Similar gradients have been
  observed for low-[$\alpha$/Fe] mono-abundance populations (MAPs) in
  the Galaxy but our results argue against interpreting them as a
  consequence of radial migration. This is because outward migration
  does not lead to thickening, implying that the maximum scale height
  of any population should reflect its value at birth. In contrast,
  Galactic MAPs have scale heights that increase monotonically
  outwards, reaching values that greatly exceed those at their
  presumed birth radii. Given the strong vertical bias affecting
  migration, a proper assessment of the importance of radial migration
  in the Galaxy should take carefully into account the strong radial
  dependence of the scale heights of the various stellar populations.
\end{abstract}


\altaffiltext{4}{Alfred P. Sloan Fellow}
\altaffiltext{5}{Senior CIfAR Fellow}

\section{Introduction}
\label{sec:intro}

Non-axisymmetric patterns in disk galaxies have long been known to
drive secular changes in the energy and angular momentum of disk stars
\citep[e.g.,][]{LB72}. Angular momentum changes are generally
accompanied by radial shifts in the average galactocentric distance of
a star. Although such changes usually lead to increased orbital
eccentricities and a gradual increase in the velocity dispersion of
the disk, \citet{Sellwood02} demonstrated, in an influential paper,
that stars can also migrate while preserving their circularities if
they are near corotation of fluctuating spiral patterns, a process
usually referred to as ``radial migration''.

This realization has led to new insights into the interpretation of a
number of properties of galaxy disks, including (i) the (weak)
correlation between age and metallicity in the solar neighborhood
\citep{E93,Haywood08,Bergemann2014}; (ii) the upturn of the mean age
in the outer parts of disks \citep{Roskar08,Bakos08}; (iii) the
formation of the Galactic thick disk \citep{SB09,Loebman11}; and (iv)
the radial dependence of the metallicity distribution function in the
Galactic disk \citep{Hayden15,Loebman16,MM16}. Although there is
consensus that migration should somehow contribute to those trends,
its actual importance in each particular case is still a matter of
debate.

One example concerns the effect of migration on the vertical structure
of the Galactic disk. Earlier work suggested that migration can
thicken a galaxy disk by pushing stars from the inner regions (where
the velocity dispersion is high) to the outskirts of the disk, where
the surface density---and hence the vertical restoring force---is
lower \citep[e.g.][]{SB09}. \citet{M12b}, on the other hand, argued
that migration has little effect on disk thickness, since outward
migrators should conserve their vertical action and cool down, a
conclusion that has been debated further by \citet{Solway2012},
\citet{Roskar2013} and \citet{VCD15}.

The vertical structure of migrators was also studied by \citet{VC14},
who noted that migrators are a heavily biased subset of stars with
preferentially low vertical velocity dispersions. This ``provenance
bias'' arises because stars with small vertical velocities spend more
time near the disk plane, and thus couple more readily to
non-axisymmetric perturbations. The resulting vertical structure of
migrators is further complicated by the fact that the velocity
dispersion of outward migrators generally decreases, whereas the
opposite applies to inward migrators.

A potential application of these ideas to the Galactic disk is
provided by recent results from the APOGEE survey. By dissecting the
Milky Way stellar disk into distinct populations according to their
$\alpha$ and Fe content, \citet[][hereafter B15]{Bovy2015} were able
to confirm the presence of two chemically-distinct populations
differing in their [$\alpha/$Fe] ratio. This distinction provides an
improved characterization of the traditional ``thick'' and ``thin''
disks of the Galaxy that does not rely on kinematics. Such purely
chemical characterization is highly preferable, since, unlike
kinematics, chemistry is a durable and stable property of a star that
should faithfully reflect its local conditions at birth \citep[see,
e.g.,][]{Navarro2011}.

An interesting property of low-[$\alpha/$Fe] stars (i.e., the
traditional ``thin disk'') is that their spatial distribution varies
smoothly and systematically with iron abundance. In particular, the
surface density of stars of a mono-abundance population (MAP) of fixed
[Fe/H] peaks at some radius and decreases both inside and outside that
radius. The ``peak radius'' increases with decreasing [Fe/H], in a
manner that defines the overall metallicity gradient of the disk.

More importantly for our current discussion, there are clear radial
trends in the thickness of each MAP: they flare outward, a result that
has been interpreted by B15 as an ``essential test of the
predictions of radial migration''. This conclusion relies on a simple
scenario where, broadly speaking, stars in each MAP share a similar
birth radius---i.e., the ``peak radius'' of their radial
distribution---and disperse through the Galaxy by the secular changes
induced by spiral patterns, flaring outward in the process.

However, given the uncertainties in our understanding of the effects of radial
migration cited earlier, this scenario seems premature. Here,
we report on results of a set of numerical simulations of isolated
stellar disks designed to test the role of radial migration in
establishing radial gradients in scale height. The simulations were
designed so that disks develop different spiral patterns, allowing us
to assess the robustness of our conclusions regarding the specific
nature of the spiral, such as number of arms and the overall strength
of the pattern.

This paper is organized as follows: Section~\ref{sec:methods}
describes the numerical simulations while our main results are
presented in Section~\ref{sec:results}. We end with a brief discussion
and a summary of our main conclusions in
Section~\ref{sec:conclusions}.

\section{NUMERICAL SIMULATIONS}
\label{sec:methods}

\subsection{Galaxy models}
The stellar disk in all of our simulations is described by an
exponential surface density profile:
$\Sigma (R)=M_{\rm d}/(2\pi R^2_{\rm d}) \exp^{(-R/R_{\rm d})}$,
with $M_{\rm d}$ the total disk mass and $R_{\rm d}$ the disk
scale-length.  The vertical distribution of disk stars follows an
isothermal sheet with a radially constant scale height, $z_{\rm  d}$.
The 3D stellar density in the disk is then:

\begin{eqnarray}
 \rho_{\rm d}(R,z) &=& \Sigma(R) \zeta(z) \nonumber \\
 &=& \frac{M_{\rm d}}{4 \pi z_{\rm d} R_{\rm d}^2} {\rm sech}^2
(z/{z_{\rm d}}) \exp{(-R/{R_{\rm d}})}
\end{eqnarray}
\noindent 

The disk mass is sampled with $N=5\times 10^{6}$ particles and is
evolved in a dark matter halo modeled as a static Hernquist profile
\citep{Hernquist1990}. One of the simulations also includes a bulge
modeled as a second Hernquist profile.

\begin{table}
\begin{center}
  \caption{Galaxy Model Parameters}
      \begin{tabular}{@{} lccr @{}}
      \hline
      
       Parameters   & \diskone{} & \disktwo{} & \diskthree{} \\ 
      \hline
      $M_{\rm d}$ $[10^{10}\; {\rm M}_{\odot}]^{a}$  &  1.91  &  4.00 & 4.00 \\
      $R_{\rm d}$ [kpc]$^{b}$                      & 3.13  &  2.84  &  2.50 \\
      $z_{\rm d}/R_{\rm d}$ $^{c}$                 & 0.10  &  0.10  & 0.15 \\
      $M_{\rm halo}$ $[10^{12}\; {\rm M}_{\odot}]^{d}$  & 0.93  &  1.00   & 9.50 \\
      $R_{\rm halo}$ [kpc]$^{e}$                      & 29.78 &  25.77  & 130.0 \\
      $M_{\rm bulge}$  $[10^{10}\; {\rm M}_{\odot}] ^{f}$ & ...    & ...      & 1.40 \\
      $R_{\rm bulge}$ [kpc]$^{g}$                      & ...    & ...      & 0.35 \\
      $f_{\rm disk}(2.2 R_{\rm d})$ $^{h}$                  & 0.28   & 0.42     & 0.48 \\ 
      \hline
      $t_0$ [Gyr]$^{i}$                              & 0.98  & 0.10   & 0.29 \\
      $t_1$ [Gyr]$^{j}$                              & 3.42 & 0.88 & 0.98 \\
      $t_2$ [Gyr]$^{k}$                              & 9.78 & 2.05 & 1.86 \\ 
      \hline 
      \hline 
     \end{tabular} 
\noindent
\\
Disk mass$^a$;  disk scale length$^b$; disk scale height$^c$;
halo mass$^d$; 
halo scale length$^{e}$; bulge mass$^{f}$ ; 
bulge scale length$^{g}$; disk mass fraction$^{h}$  within $2.2\, R_{\rm d}$; 
initial time$^{i}$;
times$^{j,k}$ when radial velocity dispersion has increased by 10-30\% respectively.  
  $  \ \ \ \ \ \ \ \ \ \ \ \ \ \ \ \ \ \ \ \ $
  \end{center}
\label{tab:params}
\end{table}

We consider three different galaxy models, labeled \diskone{} (``light
disk''), \disktwo{} (``heavy disk'') and \diskthree{} (``Milky
Way-like''). The structural parameters of each galaxy model are listed
in Table 1.

The \diskone{} galaxy is the same model as presented by
\citet{Donghia13} and analyzed in detail by \citet{VC14}. It
corresponds to a low-mass disk where the spiral patterns are recurrent
and relatively weak, but long-lived and characterized by multiple
arms.

The \disktwo{} model is a heavier-disk version of the \diskone{}
galaxy where the disk mass has been increased by a factor of two while
keeping the halo parameters basically unchanged.

Finally, the model labeled as \diskthree{} is designed to have a disk
mass and radial scalelength consistent with the Milky Way. The
addition of a bulge and the adjustment of the halo parameters yields a
flat circular velocity curve in good agreement with that of the
Galaxy; i.e., $V_{\rm circ} \sim 220 \ $km s$^{-1}$ roughly constant
in the range $0.5 < R < 14$ kpc. In this case the disk contributes
$\approx$ 70\% of the total circular speed at 2.2 scale lengths
\citep{Bovy12b}.

\begin{figure*}
  \begin{center}
    \includegraphics[width=0.98\textwidth]{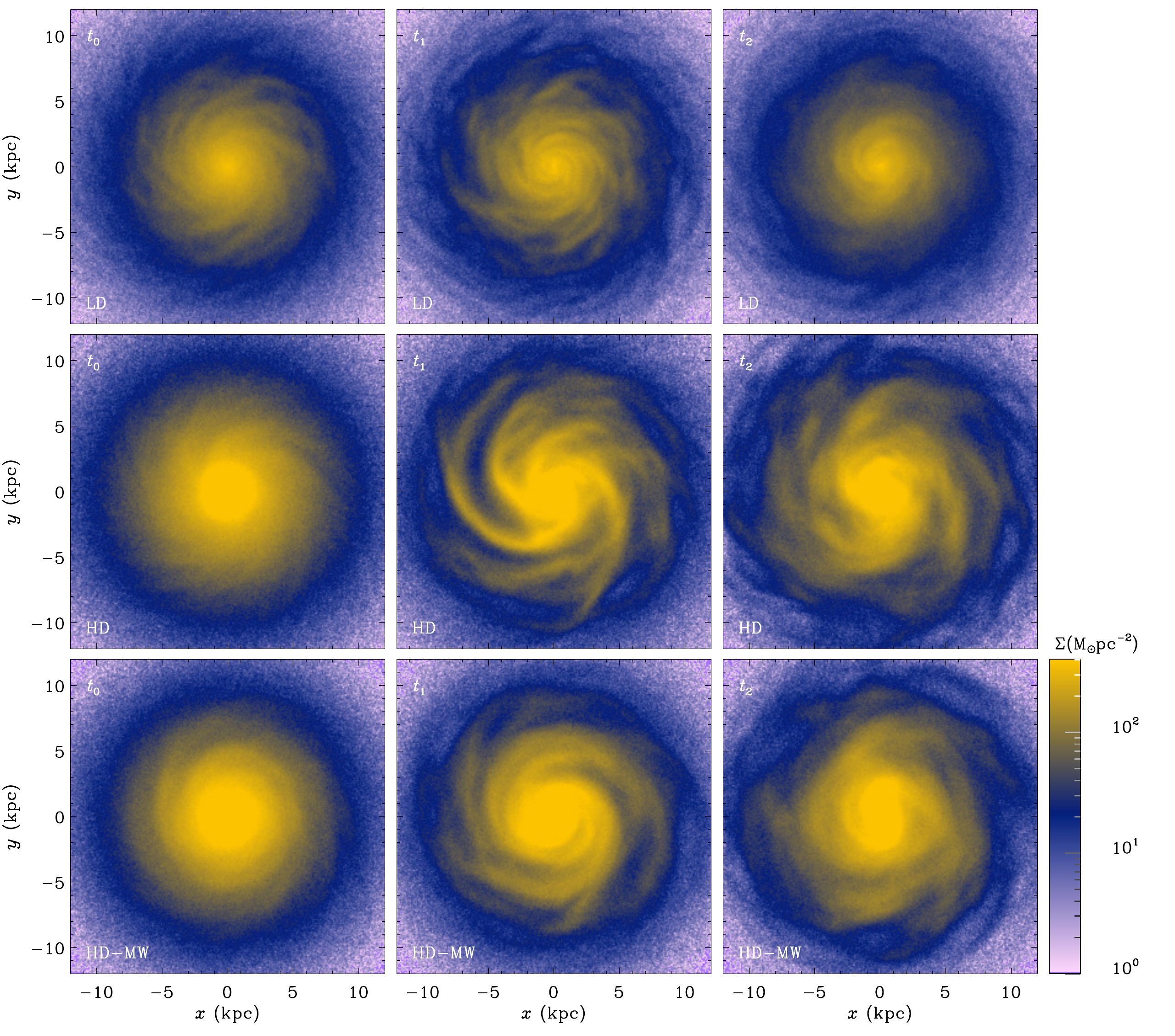}
  \end{center}
  \caption{Face-on disk projected stellar density for the three
    simulated galaxies: the low-mass disk case \diskone{} (top
    panels), and the two heavy-disk models \disktwo{} (middle panels)
    and \diskthree{} (bottom panels).  Note the different spiral
    patterns in each case. The two heavy disks eventually develop a
    strong central bar.  The structural parameters of the disks and
    characteristic times are listed in Table 1.}
  \label{fig:faceon}
\end{figure*}

\begin{figure*}
  \begin{center}
    \includegraphics[width=0.98\textwidth]{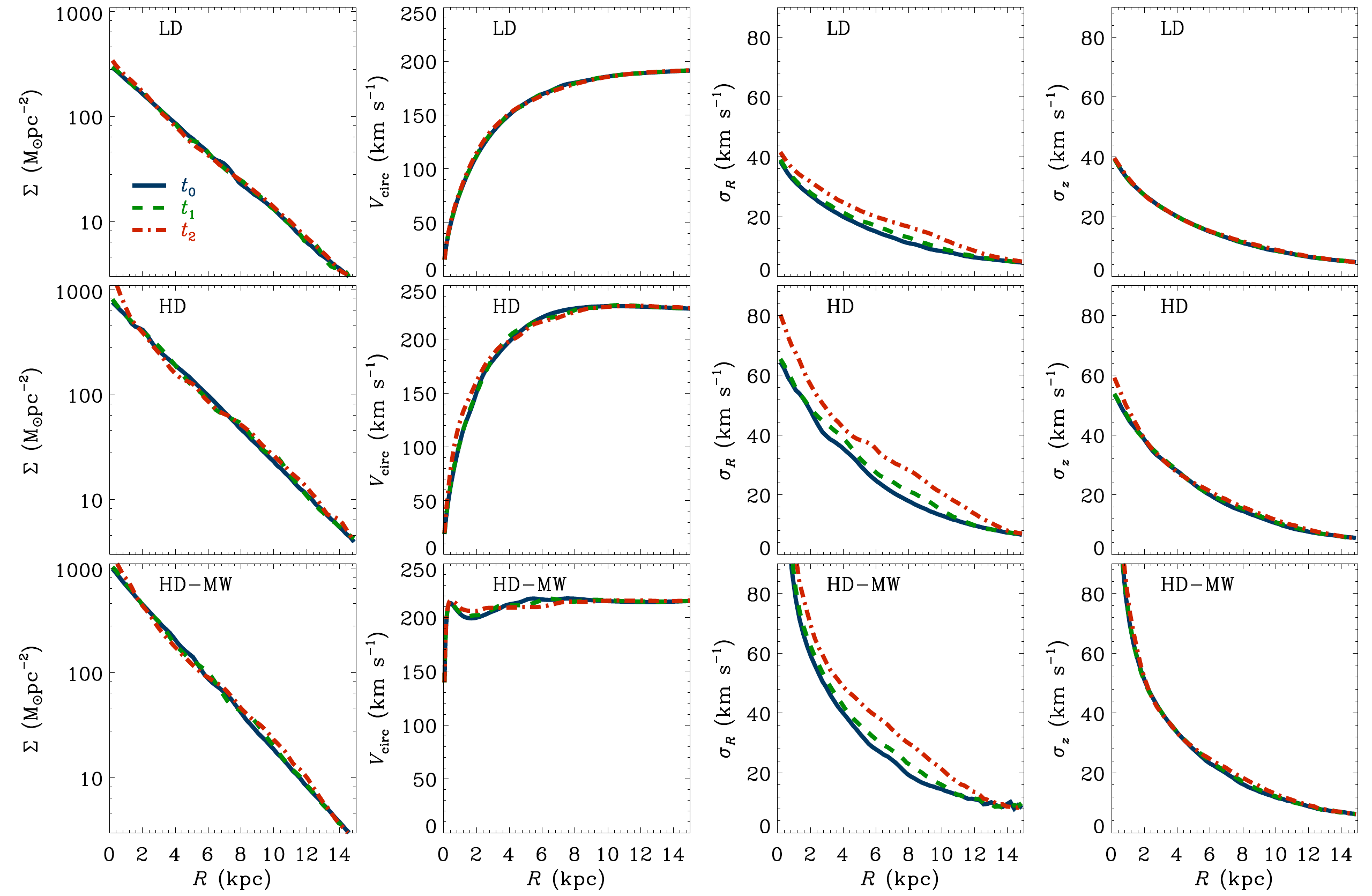}
  \end{center}
  \caption{{\it Left to right columns}: Radial profiles of stellar
    surface density, circular velocity, radial velocity dispersion and
    vertical velocity dispersion, shown at the initial time, $t_0$, an
    intermediate time, $t_1$, and the final time, $t_2$. Intermediate
    and final times are defined by the radial velocity dispersion,
    which increases by $10$ and $30\%$ since $t_0$. All times are
    listed in Table 1.}
  \label{fig:vcirc}
\end{figure*}

The radial component of the disk stellar velocity dispersion,
$\sigma_R=\sigma_\theta$, is assumed to scale proportionally to the
vertical component, $\sigma_z$: $\sigma_R^2 = f_R \sigma_z^2$, where
$\sigma_z$ is specified by the chosen scaleheight, $z_{\rm d}$
\citep{Hernquist1993}. The ratio $z_{\rm d}/R_{\rm d}$ is chosen to be
in the range $0.1$-$0.15$ for all disks.  The factor $f_R$ is chosen
so as to set a minimum value of the Toomre parameter ($Q= \sigma_R
\kappa / (3.36 \ G \Sigma)$) of unity at two disk scale lengths.


\subsection{The runs}

The simulations were carried out with the parallel TreePM code
GADGET-3. We only employ the tree-based gravity solver coupled to a
static external potential to solve for the evolution of collisionless
particles.  Pairwise particle interactions are softened with a spline
kernel \citep{HK89} on scales $h_s$, so that forces are strictly
Newtonian for particles separated by more than 2 $h_s$.  The resulting
force is roughly equivalent to traditional Plummer-softening with
scale length $\approx h_s$/2.8. For our simulations the gravitational
softening length was set to $50$ pc.

\subsection{Disk evolution}

Because of their different mass and circular velocity profiles, each
disk develops different spiral patterns over time, as shown in
Fig.~\ref{fig:faceon}. The \diskone{} model may be described as a
flocculent spiral with multiple arms, whereas the two heavy disks show
stronger patterns with fewer arms that evolve relatively quickly into
a central bar. The non-axisymmetric perturbations gradually heat the
disk in all cases, but at a much faster rate for the case of the
heavier disks than for the \diskone{} model.

In an attempt to compare the three models at comparable stages of
evolution we select three characteristic times for analysis; (i) an
initial time, $t_0$, before the spiral patterns develop and defined so
that the rms potential fluctuations on the disk plane in the radial
range $3<R/$kpc$\,<8$ is small (i.e., $\sigma_\Phi \sim 2.5 \times
10^{-4}$); (ii) an intermediate time, $t_1$, when the radial velocity
dispersion in that same radial range has increased by $10\%$; and
(iii) a final time, $t_2$, when $\sigma_R$ has increased by $30\%$
since $t_0$. We list these times (in Gyr) in Table 1. Note that it
takes many more half-mass disk rotations to heat up the \diskone{}
model (24/87) than the other two (10/26 and 13/30 for \disktwo{} and
\diskthree{}, respectively).

Fig.~\ref{fig:vcirc} shows the surface density profiles, as well as
the circular velocity and velocity dispersion profiles at those three
times for all disks. Aside from the development of a central ($\sim 1$
kpc) bar in the case of the heavy disks, the most noticeable change
from $t_0$ to $t_2$ is in the radial velocity dispersion, which, by
construction, grows by $30\%$ in all cases. Note that despite the
large and varied spiral patterns present, neither the disks thicken
noticeably nor they evolve substantially in terms of their mass
profiles.

\section{RESULTS}
\label{sec:results}

\subsection{The provenance bias of radial migration}
\label{sec:bias}

\begin{figure}
  \begin{center}
    \includegraphics[width=0.48\textwidth]{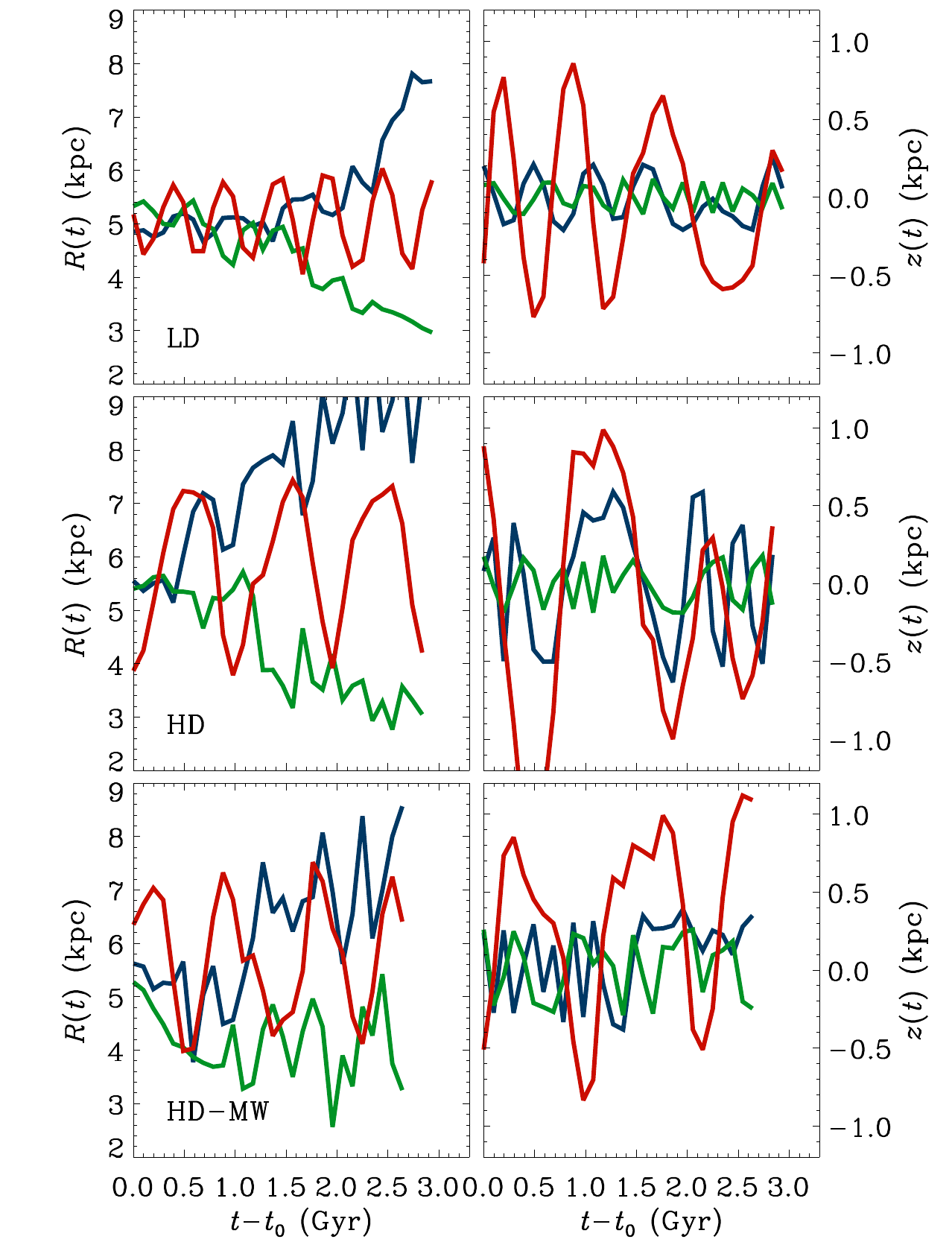}
  \end{center}
  \caption{Orbits of three particles with similar guiding center
    radius at the initial time, $R_g(t_0) = 5$ kpc, but which migrate
    outward or inward by roughly $\sim 50\%$ (blue/green curves,
    respectively.) The third (shown in red) does not show any
    substantial secular change in $R_g$.  {\it Left}: Cylindrical
    radius, $R$, as a function of time {\it Right}: Vertical
    displacement, $z$, (scale on right) as a function of time. Note
    the ``provenance bias'' of radial migrations, where extreme
    migrators are primarily particles initially on nearly circular
    orbits confined to the plane. }
  \label{fig:zmax}
\end{figure}

The non-axisymmetric patterns shown in Fig.~\ref{fig:faceon} lead to
large exchanges of angular momentum and energy between disk
particles. Since the mass profiles remain basically unchanged, a
change in specific angular momentum, $L_z$, translates directly into a
change in guiding center radius, $R_g$, defined by $L_z = R_g V_{\rm
  circ}(R_g)$, where $V_{\rm circ}^2 = \partial \Phi/\partial \ln R$
and $\Phi$ is the gravitational potential on the disk plane. The
larger the change in $R_g$ the farther a star migrates from its
original (``birth'') radius.

As discussed by \citet{VC14}, migrating stars are expected to exhibit
a heavy vertical bias: the most extreme migrators are almost
invariably those whose initial orbits do not take them far from the
disk plane. This {\it provenance bias} of migrators is not surprising:
kinematically-cold stars with modest vertical excursions spend more
time near the midplane and at the same galactocentric distance, and
are thus able to couple more effectively to non-axisymmetric
perturbations in the disk.

We illustrate this in Fig.~\ref{fig:zmax}, where we have tracked the
evolution in cylindrical radius (left panels) and height from the
plane (right panels) of three particles in each simulation, chosen so
that they share a common birth radius (i.e., $R_g(t_0) = 5$ kpc) but
which either migrate inward (green curves) or outward (blue) by
roughly $50\%$, or stay put at their original radius (red). This figure
illustrates clearly that migrating particles are preferentially those
initially close to circular orbits and with small vertical
excursions. Interestingly, we find the same qualitative result for all
three simulated disks, despite the large differences in their spiral
patterns.

The same conclusion is illustrated in Fig.~\ref{fig:bias}, where the
left panels show, for all disk stars, the fractional change in guiding
center radius, $\delta R_g = \ln R_{g}(t_2) / R_{g}(t_0)$, as a
function of birth radius, $R_g(t_0)$. The colors encode the {\it
  initial} vertical velocity dispersion of migrators, in units of the
average at the birth radius; i.e., $\sigma_z(R_g(t_0))$. Blue colors
denote kinematically-cool stars whose vertical velocity dispersion are
$\sim 20\%$ below the mean; red correspond to stars $\sim 15\%$ hotter
than the average.

The left panels of Fig.~\ref{fig:bias} demonstrate that the efficiency
of migration is a sensitive function of the initial $\sigma_z$:
extreme migrators are primarily those with small vertical motions at
birth, a conclusion that applies equally well to all three
simulations.  \citet{Grand2015} reach the same conclusion in their
study of a number of cosmological hydrodynamical simulations of Milky
Way-sized galaxies, so this finding seems robust: the provenance bias
afecting stellar migration holds regardless of the morphology of
spiral structure and strength of the non-axisymmetric perturbations.

\subsection{The vertical structure of radial migrators}

\begin{figure}
  \begin{center}
    \includegraphics[width=0.46\textwidth]{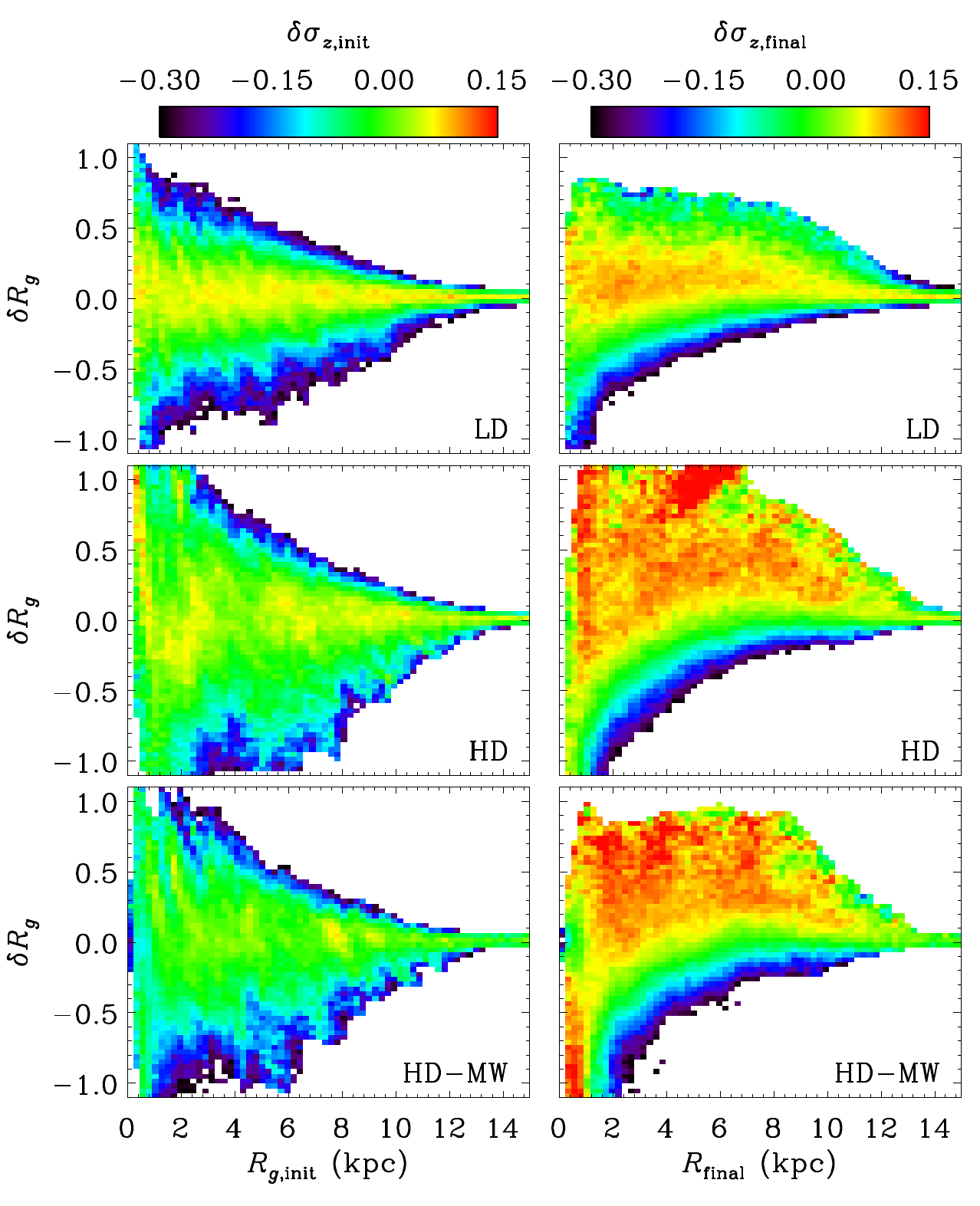}
  \end{center}
  \caption{{\it Left}: Fractional changes in guiding center radius
    ($\delta R_g = \ln R_{g}(t_2) / R_{g}(t_0)$) as a function of the
    initial ``birth radius'', $R_g(t_0)$, colored by initial vertical
    velocity dispersion, $\sigma_z(t_0)$, expressed in units of the
    average value at the birth radius. Blue indicates stars with
    initial vertical excursions smaller than the average, red the
    opposite.  {\it Right}: Same as left, but as a function of the
    final guiding center ``destination'' radius, $R_g(t_2)$, and
    colored by the final vertical velocity dispersion, scaled to the
    average at the destination radius.}
  \label{fig:bias}
\end{figure}

What is the vertical motion of migrators at their destination radii?
This is shown in the right-hand panels of Fig.~\ref{fig:bias}, which
are analogous to those on the left, but as a function of the {\it
  final} (``destination'') guiding center radius, $R_g(t_2)$, and
colored by the velocity dispersion in units of the average at the
destination radius, $\sigma_z(R_g(t_2))$. A distinction appears here
between inward and outward migrators, where inward migrators settle
into orbits much more closely confined to the plane than the average
star at at their new radii, whereas outward migrators end up with
vertical velocity dispersions close to or, in some cases, even
slightly higher than the average there.

The asymmetry between inward and outward migrators may be traced to
the strong decline with radius of $\sigma_z$ (see right-hand panels of
Fig.~\ref{fig:vcirc}) coupled with the vertical bias discussed
above. The strong radial gradient implies that particles that move out
arrive at radii where the velocity dispersion is much lower than at
their birth radius. Although they are initially biased and even cool
down a little as they move out \citep{VC14}, the gradient is so strong
that their final velocity dispersions are much closer to the average
in their new neighborhood than those that migrate inward. For the
latter, the effects of the vertical bias are amplified by the larger
velocity dispersion and higher surface densities of the inner regions:
the vertical distribution of inward migrators becomes 
thinner and thinner the further in they drift.

We should note that, although outward migrators in general do not
``heat'' the disk vertically, there are some radii where they do end
up with higher-than-average velocity dispersion. An example is
provided by the ``red band'' at $R_g(t_2)\sim 4$-$6$ kpc in the
middle-right panel of Fig.~\ref{fig:bias}. Those radii correspond to
regions in corotation with the strong bar that develops at late times
in \disktwo{}. This feature is thus caused by a bar resonance, and is
actually not present at $t_1$, before the bar develops. (Similar
patterns develop in the case of \diskthree{}, which also develops a
bar at late times.)  Indeed, no ``red band'' of vertically hot
material is seen in the case of \diskone{}, where no bar develops at
all. We should also note that these extreme migrators make up only a
small fraction of all stars at their destination radius, and their
effect on the average kinematics there is negligible, as
shown by the invariance of the $\sigma_z$ profiles shown in the
right-hand panels of Fig.~\ref{fig:vcirc}.

\begin{figure*}
  \begin{center}
    \includegraphics[width=0.98\textwidth]{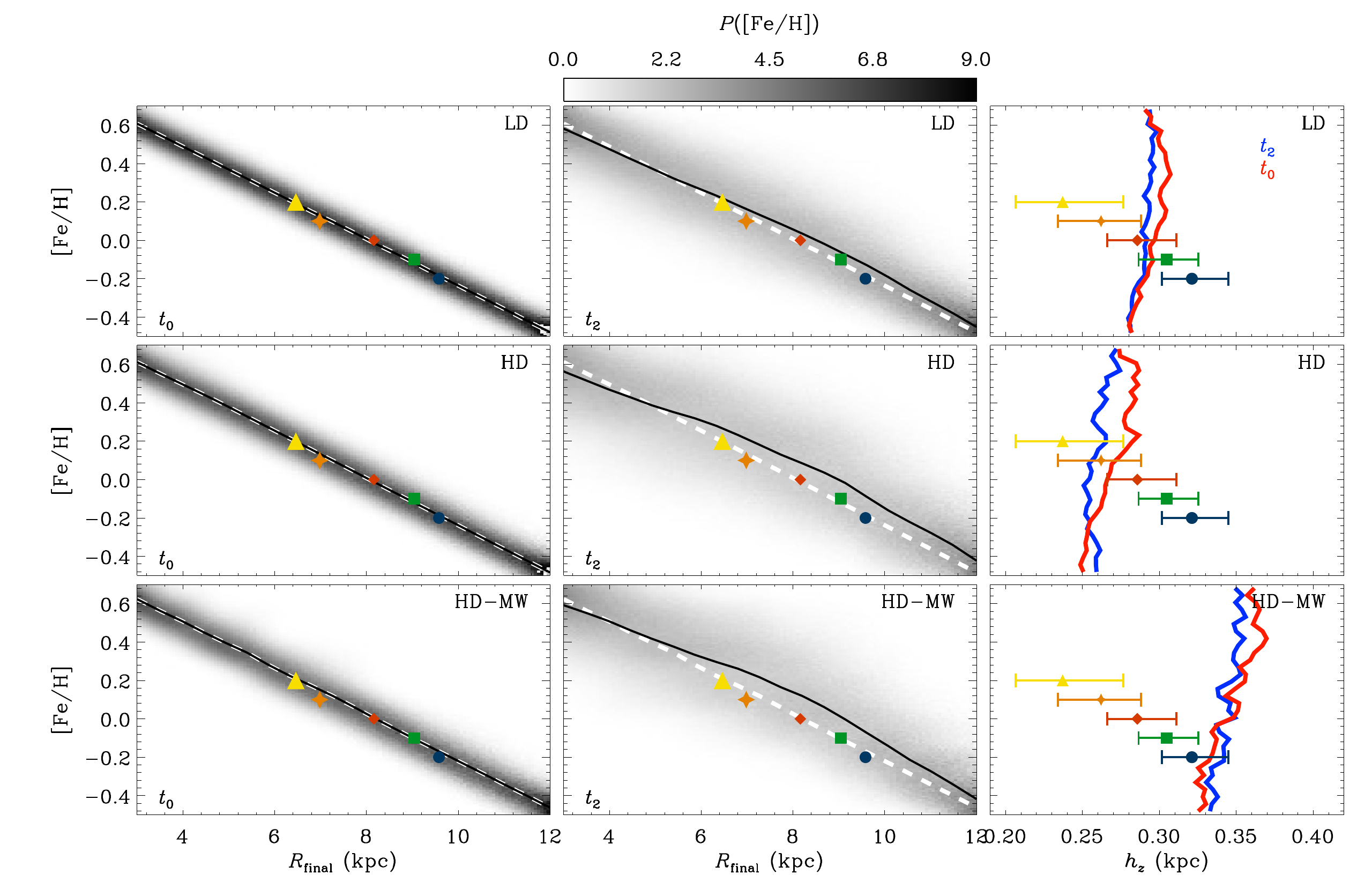}
  \end{center}
  \caption{{\it Left }: Metallicity gradient for the three simulated
    disks at $t_0$ (in grey), assumed to be $\langle {\rm
      [Fe/H]}\rangle = 0.12 \ [8\;{\rm kpc} -R_{g}(t_0)]$, with
    intrinsic width of $\sigma({\rm[Fe/H]}) = 0.04$. This is chosen to
    match the [Fe/H] dependence of the ``peak'' radius of
    mono-abundance populations in the $\alpha$-poor Galactic disk. The
    peak radius of five MAPs with [Fe/H] = -0.2, -0.1, 0.0, 0.1, 0.2,
    respectively, are shown by the colored symbols.  {\it Middle:}
    Same as left, but at $t_2$. A dashed line traces the initial
    gradient; the solid line tracks the median [Fe/H] as a function of
    radius at $t_2$. Colored symbols are unchanged from the left
    panels. {\it Right:} Exponential scale heights, $h_z$, as a
    function of [Fe/H]. Blue and red curves correspond to the initial
    and final scale height of the simulated disks, respectively. The
    initial scale height was assumed constant in our models, and
    remains basically unchanged, despite substantial radial
    migration. The Galactic disk, on the other hand, shows a strong
    trend of increasing thickness with decreasing metallicity, shown
    here at the ``peak'' radius of each MAP by the colored symbols
    with error bars. Data in all panels from B15.}
  \label{fig:metals}
\end{figure*}

\subsection{Migration and radial gradients}
\label{sec:metals}

We explore now the effects of radial migration on pre-existing radial
gradients in the disks. Migration-led mixing is expected to blur
initial trends, increasing the local dispersion and perhaps changing
the shape of the local distribution of any stellar property tightly
linked with its birth radius. One example of this is provided by the
metallicity distribution function in the Galactic plane, which changes
skewness inside/outside the solar circle, a feature probably caused by
migration \citep{Hayden15}.

In order to address such issues, we tag stars, at the initial time,
$t=t_0$, according to the following relation:

\begin{equation}\label{eq:metals}
  \langle {\rm [Fe/H]}\rangle = 0.12 \ [8\;{\rm kpc}
  -R_{g}(t_0)],
\end{equation} 
with a Gaussian dispersion of $\sigma({\rm[Fe/H]}) = 0.04$.

The slope of this equation is chosen so as to reproduce the radial
dependence of the MAP peak radius reported by B15 for the
Milky Way's $\alpha$-poor disk.  We show this in the left panels of
Fig.~\ref{fig:metals}, which display the initial metallicity gradient
of each disk according to Eq.~\eqref{eq:metals}, compared with the
``peak radii'' of five mono-abundance populations ([Fe/H]$=-0.2, -0.1,
0.0, 0.1, 0.2$, respectively). By construction, our initial disks have
metallicities that match, at their birth radii, those of MAPs in the
$\alpha$-poor (``thin'') disk of the Galaxy.

How would the metallicity gradient evolve as a consequence of radial
migration?  This is shown in the middle panels of
Fig.~\ref{fig:metals}, where we show the final gradient, computed at
$t_2$. Although migration significantly blurs the initial trend,
increasing the dispersion at given radius, it does not affect the
slope of the overall gradient much. Interestingly, it does not change
the scale-height profile of the disk either, as may be seen in the
right-hand panels of Fig.~\ref{fig:metals}. Here the solid blue and
red curves running vertically indicate the exponential scale-height of
stars at $t_0$ and $t_2$, respectively, as a function of
[Fe/H]. Clearly, migration has had little effect on the overall disk
scale height, measured either at fixed radius, or at fixed
metallicity.

\subsection{Vertical structure of radial migrators}
\label{sec:flares}

The right-hand panels of Fig.~\ref{fig:metals} also illustrate one of
the main differences between our models and the Galactic disk. Our
models assume a {\it constant} initial scale height, which is
basically unchanged by migration, and cannot therefore match, through
the simple tagging proposed by Eq.~\eqref{eq:metals}, the marked
increase in thickness with decreasing metallicity observed in the
Galaxy (colored symbols with error bars). The inability of radial
migration to thicken the disk strongly suggests that the observed
increase in Galactic disk thickness with decreasing metallicity (and
increasing radius) is probably an intrinsic feature of the disk, and
not a consequence of radial migration.

As mentioned in Sec.~\ref{sec:intro}, one interesting property of
Galactic mono-abundance populations is the presence of radial
gradients in their vertical structure. These affect primarily MAPs in
the low-[$\alpha$/Fe] (``thin'') disk, whose scale heights increase
monotonically with radius (B15). The gradients are such that, at fixed
[Fe/H], thin-disk stars are thinner inside their ``peak radius'', and
thicker outside. For example, Sun-like stars ([Fe/H]$=0$) ``peak'' at
$\sim 8$ kpc, where their exponential scale-height is $h_z \sim 300$
pc: their scale height is a factor of $\sim 1.6$ thinner at $R\sim 4$
kpc, and $\sim 1.6$ times thicker at $R\sim 14$ kpc. Could this be the
result of radial migration?

\begin{figure}
  \begin{center}
    \includegraphics[width = 0.48\textwidth]{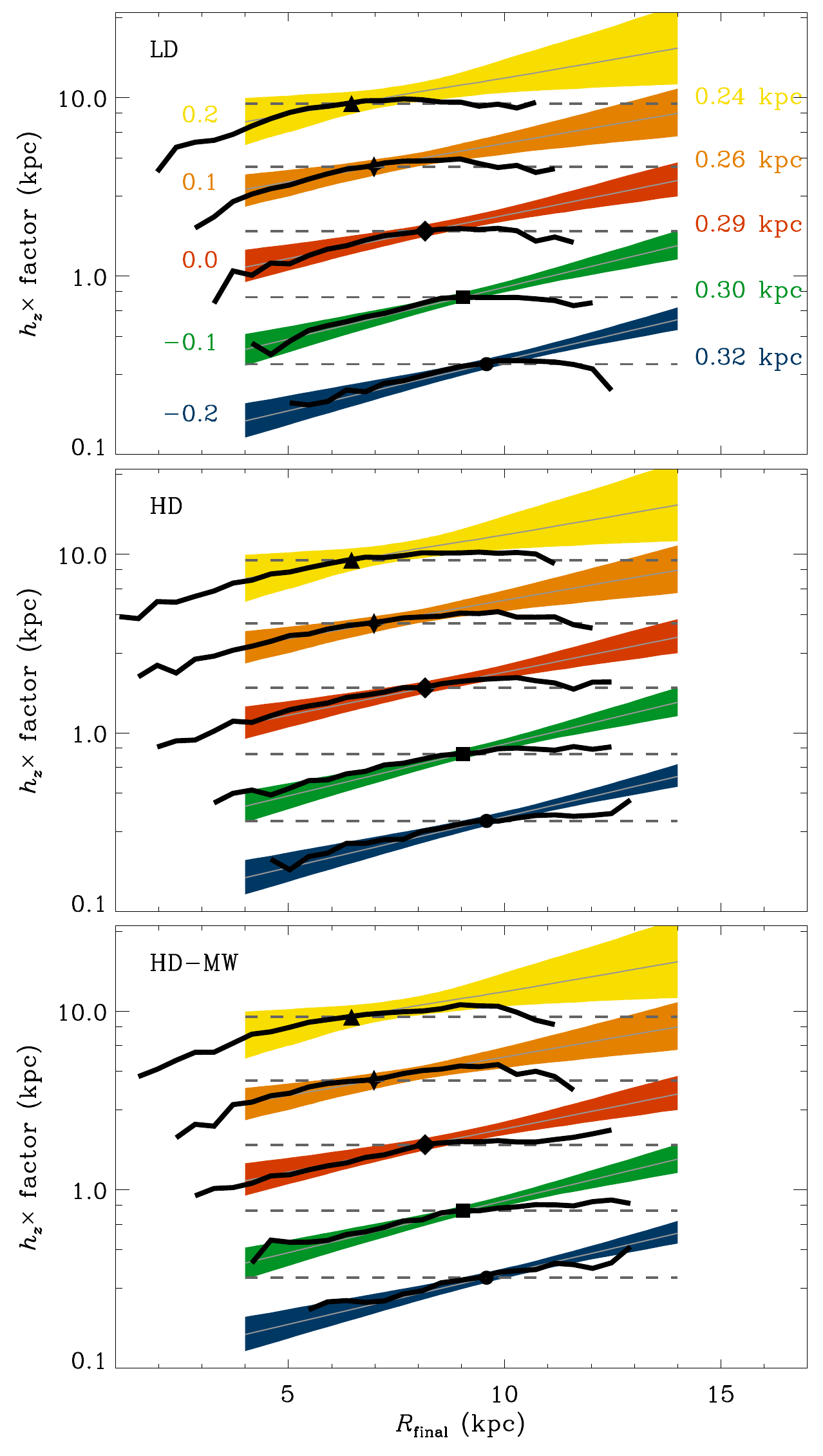}
  \end{center}
  \caption{Radial dependence of the scale height of mono-abundance
    populations with [Fe/H] = -0.2, -0.1, 0.0, 0.1, 0.2,
    respectively. Black curves indicate results for the simulated
    disks at $t_2$, colored bands show the results for the
    $\alpha$-poor Galactic disk from B15. Results for
    each MAP have been shifted vertically for clarity. The height of
    each simulated MAP has also been rescaled to match the observed
    value of the corresponding Galactic MAP at its ``peak''
    radius. The latter are indicated by heavy symbols. Note that
    migrating stars thin down as they spread inward, but do {\it not}
    thicken up as they move outwards. The observed radial trend is
    therefore unlikely to be caused by radial migration, particularly
    the outer ``flaring'' of each MAP.}
  \label{fig:thick}
\end{figure}

We address this question in Fig.~\ref{fig:thick}, where the thick
solid curves display the radial profile of the scale height, $h_z(R)$,
for $5$ MAPs with metallicities selected between $-0.2 <$ [Fe/H]$<
0.2$ in bins $0.1$ dex apart. To compare with the results reported by
B15, we rescale each of the scale heights of the simulated MAPs to
coincide with that observed in the Galaxy at their corresponding
``peak'' radius (indicated by the symbols). The profile of each MAP is
further shifted vertically by an arbitrary amount for clarity. The
color bands show the scale-height profile of the MAPs observed in the
Milky Way (B15).

The main conclusion of this comparison is that radial migration, at
least in our simulations, does not match the observed trends in
the Galaxy. Migrators do thin down as they spread inward, but they do
{\it not} thicken up as they move outward. As a result, although the
vertical trend of a given MAP {\it inside} the peak radius agrees relatively
well with the observed trend, {\it outside} that radius the simulated scale
heights are systematically below those observed.

We emphasize that our results do not rule out that migration might
play {\it some} role in the vertical trend, but, if it did, it would
be responsible for the ``thinning down'' of a population inside its
peak radius, rather than for its flaring outside of it, as is usually
envisioned in scenarios where the trend is driven by migration
(Sec.~\ref{sec:intro}). A further difficulty of such scenario concerns
the strong sensitivity of radial migration to the initial height of
the population, and the fact that it would take longer for stars born
in the outskirts of the Galaxy to move inward, given their long
orbital times. 

The latter is a strong constraint, given the strong dependence of
migration efficiency with radius seen in Fig.~\ref{fig:bias}: whereas
changes in radius of $100\%$ are not unusual at $\sim 4$-$5$ kpc, they
do not exceed $10\%$ at $R\sim 12$ kpc, mainly as a result of the
increase in orbital time with radius.  It is unclear then how stars in
the Galaxy with, say, [Fe/H]$\sim -0.5$---which would be born
primarily at $R\sim 12$ kpc with a thickness of nearly $1$ kpc
(B15)---would be able to spread efficiently around the
Galaxy through radial migration, given their  long orbital
times and large thickness.

We end by noting that our models do not match the vertical structure
of the Galaxy as a function of radius, neither at the beginning nor at
the end of the simulations, so our conclusions are probably not the
final word on this topic. Nevertheless, our discussion should serve to
emphasize the critical role of provenance bias on radial migration, and
the importance of carefully modeling the radial and vertical structure of the
various Galactic populations when assessing the effects of migration.

\section{Conclusions}
\label{sec:conclusions}

We have used $N$-body models of isolated disk galaxies with realistic
mass and circular velocity profiles to study the effects of radial
migration. We focus on the vertical bias (which we term ``provenance
bias'') favoring the migration of kinematically-cold stars on nearly
circular orbits confined to the disk \citep{VC14}. The models contrast
the effects of migration in a low-mass disk with weak, slowly-evolving
multi-armed spirals with those in heavy disks where the spiral patterns
are stronger and with fewer arms, evolving quickly into a central bar.

Our main conclusions may be summarized as follows:

\begin{itemize}

\item Provenance bias is present in all of our simulations, regardless
  of the nature of the spiral pattern. This bias implies that the
  efficiency of migration will depend sensitively on the thickness of
  a particular stellar population, a feature that must be taken
  carefully into account when modeling the effects of radial
  migration.

\item Provenance bias has a strong effect on the vertical structure of
  stars that have migrated away from their initial ``birth''
  radius. Migrators are generally a kinematically colder subset whose
  vertical velocity dispersion typically drops as they move out or
  increase as they move in. Their final structure is therefore
  heavily dependent on the radial gradient of the vertical structure
  of the disk.

\item In our models, which feature a constant scale height and a
  strong radial gradient in $\sigma_z$, inward migrators become more
  and more heavily biased relative to the average population at their
  destination radius. Outward migrators, on the other hand, move to
  regions of lower $\sigma_z$ and become a closer match to the average
  population at their final radii. 

\item In general, radial migrators thin down as they move in, but do
  not substantially thicken up as they move out, at least in disks
  like the ones we consider here. Radial migration alone thus does not
  provide a natural explanation for the monotonic increase with radius
  of the scale height of mono-abundance populations in the
  $\alpha$-poor Galactic disk reported by the APOGEE survey (B15).

\end {itemize}

Our results demonstrate that ``provenance bias'' plays a crucial role in
the final vertical structure of stars that have migrated as a
consequence of internal processes such as internally-driven spiral
patterns. Such migration does not lead, in general, to thicker disks,
suggesting that the strong and monotonic increase of scale height seen
in the $\alpha$-poor Galactic disk has a different origin. Either the
trend is inherent to the disk, or it was driven by external
perturbations, such as accretion events and collisions with dark
matter substructures. 

A more definitive assessment of the importance of radial migration in
the Galaxy will likely require cosmological models that are able to
reproduce in detail the observed trends in the radial and vertical
distributions of the various mono-abundance populations that make up
the disk. Encouraging first steps have already been taken \citep[see,
e.g.,][and references therein]{Grand2015}, but a full understanding of
the relative importance of internal and external mechanisms in shaping the
Galactic disk seems still beyond reach.

\vskip 3em

This research has been partially funded by ATP NASA Grant No
NNX144AP53 and by the National Science Foundation under grants NSF
AST-1211258 and NSF PHY11-25915. CVC and JFN acknowledge the
hospitality of the Kavli Institute for Theoretical Physics at the
University of California, Santa Barbara.  ED gratefully acknowledges
the support of the Alfred P. Sloan Foundation.  Simulations have been
run on the High Performance Computing cluster provided by the Advanced
Computing Infrastructure (ACI) and Center for High Throughput
Computing (CHTC) at the University of Wisconsin. We are grateful to
Rob Grand for his very constructive comments and to Jo Bovy for kindly
granting us access to his results.

\bibliographystyle{apj} 
\bibliography{refsnew}

\end{document}